\documentclass[twocolumn,english,amsart,showpacs,preprintnumbers,amsmath,amssymb,floatfix]{revtex4-1}
\usepackage{tikz,xcolor}
\usepackage[colorlinks = true,  
linkcolor = blue,
urlcolor  = blue,
citecolor = blue,
anchorcolor = blue]{hyperref} 

\definecolor{lime}{HTML}{A6CE39}
\DeclareRobustCommand{\orcidicon}{%
	\begin{tikzpicture}
	\draw[lime, fill=lime] (0,0) 
	circle [radius=0.16] 
	node[white] {{\fontfamily{qag}\selectfont \tiny ID}};
	\draw[white, fill=white] (-0.0625,0.095) 
	circle [radius=0.007];
	\end{tikzpicture}
	\hspace{-2mm}
}

\foreach \x in {A, ..., Z}{%
	\expandafter\xdef\csname orcid\x\endcsname{\noexpand\href{https://orcid.org/\csname orcidauthor\x\endcsname}{\noexpand\orcidicon}}
}

\usepackage[T1]{fontenc}
\usepackage[latin9]{inputenc}
\usepackage{color}
\usepackage{array}
\usepackage{amstext}
\usepackage{graphicx}
\usepackage{esint}
\usepackage{rotating}
\usepackage{appendix}
\usepackage{float}

\usepackage[font=small,labelfont=bf]{caption}
\usepackage{xcolor}
\usepackage{ulem}

\makeatletter

\@ifundefined{textcolor}{}
{%
	\definecolor{BLACK}{gray}{0}
	\definecolor{WHITE}{gray}{1}
	\definecolor{RED}{rgb}{1,0,0}
	\definecolor{GREEN}{rgb}{0,1,0}
	\definecolor{BLUE}{rgb}{0,0,1}
	\definecolor{CYAN}{cmyk}{1,0,0,0}
	\definecolor{MAGENTA}{cmyk}{0,1,0,0}
	\definecolor{YELLOW}{cmyk}{0,0,1,0}
}

\@ifundefined{definecolor}
{\usepackage{color}}{}
\@ifundefined{definecolor}
{\usepackage{color}}{}
\makeatother
\usepackage{babel}

\begin{document}
	
	%%%%%%%%%%%%%%%%%%%%%%%%%%%%%%

%\preprint{APS/123-QED}

\title{Sensitivity of strange quark polarization to flavor $SU(3)$ symmetry breaking}

% \altaffiliation[Also at ]{}%Lines break automatically or can be forced with \\

\author{Ali Khorramian}
% \altaffiliation[Also at ]{}%Lines break automatically or can be forced with \\
\email{Khorramiana@semnan.ac.ir}
\affiliation{Faculty of Physics, Semnan University, 35131-19111, Semnan, Iran 
}%

\author{Elliot Leader}
\email{e.leader@imperial.ac.uk}
\affiliation{Imperial College,
Prince Consort Road, London SW7 2BW, England
}%

\author{Dimiter B. Stamenov}
\email{stamenov@inrne.bas.bg}
\affiliation{Institute for Nuclear Research and Nuclear Energy,
Bulgarian Academy of Sciences,\\
Boulevard Tsarigradsko Chaussee 72, Sofia 1784, Bulgaria}%

\author{Amir Shabanpour}
% \altaffiliation[Also at ]{}%Lines break automatically or can be forced with \\
\email{amir.shabanpourph@semnan.ac.ir}
\affiliation{Faculty of Physics, Semnan University, 35131 {}-19111, Semnan, Iran
}%


\begin{abstract}
The polarized strange quark puzzle concerns the fact that the polarized 
strange quark density extracted from polarized \textit{inclusive} deep 
inelastic scattering data is significantly negative, whereas it is zero 
or slightly positive when extracted from a combined analysis of polarized 
\textit{semi-inclusive} and inclusive deep inelastic data. $SU(3)$ flavor 
symmetry, which, it is generally accepted,  is not an exact symmetry, 
plays an important role in the inclusive analysis, and all the extracted 
polarized quark densities depend, to some extent,  on the level of symmetry 
breaking introduced. But by far the most sensitive to the breaking is the   
strange quark density. In this paper we present a NLO QCD analysis of 
the world data on polarized inclusive DIS data on protons, neutrons and deuterons, including  the final JLAB CLAS/EG1b data on the proton  and deuteron, and study the sensitivity of the strange quark polarization 
to the breaking of flavor $SU(3)$.

\end{abstract}

\maketitle

\section{Introduction}

Our knowledge of the internal partonic spin structure of the nucleon 
comes mainly from the polarized inclusive and semi-inclusive 
deep inelastic scattering (DIS) of leptons on nucleons. While 
the inclusive DIS processes, in the lack of neutrino reactions on a polarized target at present, determine only the sum of quark 
and antiquark polarized parton density functions (PPDFs), 
($\Delta q(x) + \Delta \bar{q}(x)$), the polarized semi-inclusive DIS 
data could provide information about the individual polarized quark 
and antiquark densities, if the fragmentation functions (FFs) are 
well determined.

All QCD analyses of the polarized \textit{inclusive} DIS data, in which the flavor $SU(3)$ symmetry is not broken, have produced \textit{negative} values for the polarized strange quark density, ($\Delta s(x,Q^2) + \Delta \bar{s}(x,Q^2)$) (see for instance (\cite{LSS07}-\cite{LSS15})). 
Note that in the majority of these analyses simple input parametrizations for the polarized strange quark density, which do not permit a sign change of the density, have been used.

It was shown \cite {LSS15}, however, that even allowing in the parametrization of the polarized strange quark density for a possible sign change, in the presence of the then available more precise data, the inclusive analysis still yielded significantly negative values for the polarized strange quark density, except for negligible positive values in the region $x > 0.3$.

On the other hand, the strange quark density obtained from combined 
QCD analyses of inclusive and semi-inclusive polarized deep inelastic 
scattering data (\cite{DSSV}-\cite{AKS}) turns out to be positive in 
the $x$ range $0.02 < x < 0.2$ and negative for $x < 0.02$.
This disagreement between the inclusive and semi-inclusive analyses is 
known as the strange quark polarization puzzle. 

It was shown \cite{LSS11} and understood that in the presence of the polarized semi-inclusive data the polarized strange quark density is very
sensitive to the kaon fragmentation functions used in the analysis, and more generally, that to obtain correct values for the polarized individual quark and antiquark densities it is crucial that the fragmentation functions to be reliably determined.

It is important to mention that in the QCD analyses of the polarised  inclusive DIS data {\it alone}, the $SU(3)$ symmetric value 3F-D for the nonsinglet axial charge $a_8$ (with parameters F and D determined from the hyperon $\beta$ decays analysis, see for instance \cite{Cabibo}) is usually used, but it is of interest to know to what extent the $SU(3)$ symmetry is believed to be broken. There is a growing precision of the measurements of magnetic moments, $g_A/g_V$ ratios and rates in hyperon $\beta$ decays, and different theoretical models (\cite {noSU3models} and more recently \cite {Bass-Thomas}) have been used to study how large any flavour $SU(3)$ symmetry breaking should be in order to describe well these new data, and consequently how big could be the deviation of $a_8$ from its $SU(3)$ symmetric value. In this connection we would like to mention the QCD analysis \cite {NNPDF} of the polarised inclusive DIS data, where for the uncertainty of the $SU(3)$ symmetry value of $a_8$ has been used a value approximately six times bigger than that obtained from the phenomenological analysis of the data on hyperon $\beta$ decays. This choice of the uncertainty actually means that the SU(3) symmetry is up to 30 \% broken. 

In 2000 the first NLO QCD analysis \cite {LSS2000} of the polarized inclusive DIS data was performed in order specifically to study the sensitivity of the polarized PDFs and their first moments to the $SU(3)$ symmetry breaking effects which were taken into account in model independent way. The main result was that most sensitive to the change of the value 
of $a_8$ is the polarized strange quark density, but it remains negative 
for all $x$.

In this paper we present a NLO QCD analysis of the world data set on the nucleon spin structure functions $g_1^N$ ($N=p, n, d$) including the final JLAB CLAS/EG1b data on the proton \cite {EG1bp} and deuteron \cite {EG1bd} spin structure functions. The aim of our analysis is to further study the sensitivity of the polarized PDFs, and especially the polarized strange quark density, to the change of the value of the nonsinglet axial charge $a_8$ due to $SU(3)$ symmetric breaking effects, now that we have much more data and with higher accuracy and wider kinematic range than those 
available in 2000.  

Unlike to the QCD analyses of the polarized inclusive DIS data usually performed the following changes are made:

~~~(i) We use now input parametrizations for the sum of quark and 
antiquark polarized PDFs $\Delta q(x) + \Delta \bar{q}(x)$ instead of the valence and sea quarks densities,  because as was
stressed above only the sums $\Delta q(x,Q^2) + \Delta
\bar{q}(x,Q^2)$ can be extracted from the data. 

~~(ii) We do not make any assumptions about the polarized light
sea quark densities $\Delta \bar{u}(x)$ and $\Delta \bar{d}(x)$
which have been used in almost all previous analyses, because  
such assumptions cannot be directly tested. Note here that in contrast to the light sea quark densities, the total strange quark density 
$(\Delta s + \Delta \bar{s})(x,Q^2)$ can be well determined from the inclusive data if they are sufficiently precise \cite{LSS1998}.

\section{QCD Framework for Inclusive Polarized DIS}

One of the features of polarized DIS is that more than half of the
present data are at moderate $Q^2$ and hadronic final state mass squared $W^2$ ($Q^2 \sim 1-5~\rm GeV^2,~4~\rm GeV^2 < W^2 < 10~\rm GeV^2$), or in the so-called {\it preasymptotic} region. 
This is especially the case for the
very precise experiments performed at the Jefferson Laboratory.
So, in contrast to the unpolarized case this region cannot be
excluded from the analysis. As was shown in \cite{LSS07},
to confront correctly the QCD
predictions to the experimental data including the preasymptotic
region, the {\it non-perturbative} higher twist (powers in
$1/Q^2$) corrections to the nucleon spin structure functions have
to be taken into account too.

In QCD the spin structure function $g_1$ has the following form for $Q^2 >> \Lambda_{QCD}^2$ (the nucleon target label ``$N$" is not shown):
\begin{equation}
g_1(x, Q^2) = g_1(x, Q^2)_{\rm LT} + g_1(x, Q^2)_{\rm HT}~,
\label{g1QCD}
\end{equation}
where ``LT" denotes the leading twist ($\tau=2$) contribution to
$g_1$, while ``HT" denotes the contribution to $g_1$ arising from
QCD operators of higher twist, namely $\tau \geq 3$.  
\begin{eqnarray}
g_1(x, Q^2)_{\rm LT}&=& g_1(x, Q^2)_{\rm pQCD}  + h^{\rm TMC}(x,
Q^2)/Q^2\nonumber\\
&&+ {\cal O}(M^4/Q^4)~, \label{g1LT}
\end{eqnarray}
where $g_1(x, Q^2)_{\rm pQCD}$ is the well known (logarithmic in
$Q^2$) NLO pQCD contribution
\begin{eqnarray}
g_1(x,Q^2)_{\rm pQCD}&=&\frac{1}{2}\sum _{q} ^{n_f}e_{q}^2
[(\Delta q +\Delta\bar{q})\otimes (1 +
\frac{\alpha_s(Q^2)}{2\pi}\delta C_q)
\nonumber\\
&&+\frac{\alpha_s(Q^2)}{2\pi}\Delta G\otimes \frac{\delta C_G}
{n_f}]~, \label{g1partons}
\end{eqnarray}
and $h^{\rm TMC}(x, Q^2)$ are the exactly calculable kinematic target mass
corrections \cite{TMC}, which, being purely kinematic, effectively belong 
to the LT term.
In Eq. (\ref{g1partons}), $\Delta q(x,Q^2), \Delta\bar{q}(x,Q^2)$
and $\Delta G(x,Q^2)$ are quark, anti-quark and gluon polarized
densities in the proton, which evolve in $Q^2$ according to the
spin-dependent NLO DGLAP equations. $\delta C(x)_{q,G}$ are the
NLO spin-dependent Wilson coefficient functions calculated in 
$\rm \overline{MS}$ scheme and the symbol $\otimes$ denotes the usual convolution in Bjorken $x$ space.
$n_f$ is the number of active flavors ($n_f=3$ in our analysis).

In addition to the LT contribution, the dynamical higher twist
effects
\begin{equation}
g_1(x, Q^2)_{\rm HT}= h(x, Q^2)/Q^2 + {\cal O}(\Lambda^4/Q^4)~,
\label{HTQCD}
\end{equation}
must be taken into account at low $Q^2$. The latter are
non-perturbative effects and cannot be calculated in a model
independent way. That is why we prefer to extract them directly
from the experimental data. Note also, that in our
analysis the logarithmic $Q^2$ dependence of $h(x, Q^2)$ in Eq.
(\ref{HTQCD}), which is not known in QCD, is neglected. Compared
to the principal $1/Q^2$ dependence it is expected to be small and
the accuracy of the present data does not allow its determination.
Therefore, the extracted from the data values of the parameters 
$h^N(x_i)$ ($N=p,~n;~i=$1,2,..5) correspond to the mean $Q^2$ for each $x_i$-bin (see Fig. 6 and the discussion there). 

In our analysis of the inclusive DIS data the inverse Mellin transformation method has been used to calculate the spin structure function 
$g_1(x,Q^2)_{\rm LT}$ from its moments taking into account the first order 
in ${\cal O}(M^2/Q^2)$ TMC. For the numerical calculations the Pegasus routines \cite{num_calc} have been used. 

As was mentioned in the Introduction, we are using now input parametrizations at $Q_0^2=1$ GeV$^2$ for the sum of quark and anti-quark polarized parton densities instead of the valence and sea quarks densities:
\begin{eqnarray}
\nonumber
x(\Delta u+\Delta \bar{u})(x,Q^2_0)&=&A_{u_{+}}x^
{\alpha_{u_{+}}}
(1-x)^{\beta_{u_{+}}}\\
\nonumber
&&(1+\epsilon_{u_{+}}{\sqrt{x}}+
\gamma_{u_{+}}x),\\[2mm]
\nonumber
x(\Delta d+\Delta \bar{d})(x,Q^2_0)&=&A_{d_{+}}x^
{\alpha_{d_{+}}}
(1-x)^{\beta_{d_{+}}}(1+\gamma_{d_{+}}x),\\[2mm]
\nonumber
x(\Delta s+\Delta \bar{s})(x,Q^2_0)&=&A_{s_{+}}x^
{\alpha_{s_{+}}}
(1-x)^{\beta_{s_{+}}}(1+\gamma_{s_{+}}x),\\[2mm]
x\Delta G(x,Q^2_0)&=&A_{G}x^{\alpha_{G}}
(1-x)^{\beta_G}(1+\gamma_{G}x),
\label{input_PDFs}
\end{eqnarray}
and do {\it not} use any assumptions about the light sea quark
densities $\Delta \bar{u}$ and  $\Delta \bar{d}$. In
(\ref{input_PDFs}) the notation $q_{+}=q+\bar{q}$ is used for
$q=u,d,s$.

Usually the set of free parameters
in (\ref{input_PDFs}) is reduced by the well known sum rules
\begin{equation}
a_3=g_{A}=\rm {F+D}=1.270~\pm~0.003
\label{ga}
\end{equation}
\begin{equation}
a_8=3\rm {F-D}=0.586~\pm~0.031
\label{3FD}
\end{equation}
where $a_3$ and $a_8$ are non-singlet combinations of the first
moments of the polarized parton densities corresponding to
$3^{\rm rd}$ and $8^{\rm th}$ components of the axial vector
Cabibbo current
\begin{eqnarray}
%\nonumber
a_3&=&(\Delta u+\Delta\bar{u})(Q^2) - (\Delta
d+\Delta\bar{d})(Q^2),\\[2mm]
\nonumber
a_8&=&(\Delta u +\Delta\bar{u})(Q^2) + (\Delta d +
\Delta\bar{d})(Q^2)\\
&-&2(\Delta s+\Delta\bar{s})(Q^2),
\end{eqnarray}
and the values of parameters F and D are determined from the 
$SU(3)$ flavor symmetry analysis of the hyperon $\beta$ decays,
and slightly change over the years due to the improvement of the precision of the experiments. The experimental values for $g_A$ in (\ref{ga}) and $a_8$ in (\ref{3FD}), used in our study, are presented in \cite{PDG} and \cite{HERMES}, respectively.  

The sum rule (\ref{ga}) reflects isospin $SU(2)$ symmetry, whereas
(\ref{3FD}) is a consequence of the $SU(3)_f$ flavor symmetry
treatment of the hyperon $\beta$-decays. So, using the constraints
(\ref{ga}) and (\ref{3FD}) the parameters $A_{u+\bar{u}}$ and
$A_{d+\bar{d}}$ in (\ref{input_PDFs}) are determined as functions
of the parameters connected with $(\Delta u+\Delta
\bar{u}),~(\Delta d+\Delta \bar{d})$ and $(\Delta s+\Delta \bar{s})$.

The large $x$ behaviour of the polarized PDFs is mainly controled by
the positivity constraints \cite{LSS10}. The only difference is 
that now we are using for the unpolarized NLO PDFs the MMHT'14 set of parton densities \cite{MMHT14} instead the MRST'02 one. We have found that the positivity condition for the polarized strange quarks and gluons is guaranteed, if for the values of the parameters $\beta_{s+\bar{s}}$ and $\beta_G$, which control their large $x$ behavior, the values 9.0 and 5.0 are used, respectively. 

The rest of the parameters $\{A_i, \alpha_i, \beta_i, \epsilon_i,
\gamma_i\}$, as well as the unknown higher twist corrections
$h^N(x_i)/Q^2$ to the spin structure functions $g_1^N(x,Q^2),~(N=p,
n)$ have been determined simultaneously from the best fit to the
DIS data. Note that the $\sqrt{x}$ term has been used only in the
parametrization for the $(\Delta u+\Delta\bar{u})$ density,
because the parameters $\epsilon_{i}$ in front of it for the other
polarized densities can not be determined from the fit, and do not
help to improve it. The parameter $\gamma_G$ was fixed to zero because 
the accuracy of the present data do not also allow its determination. 
Concerning the parameter $\gamma_{s_{+}}$ see the discussion below.

The method used to extract simultaneously the polarized parton 
densities and higher twist corrections from the data is described 
in \cite{LSS_HT}. 

In polarized DIS the $Q^2$ range and the accuracy of the data are
much smaller than that in the unpolarized case. That is why, in
all calculations we have kept fixed the value of the strong coupling constant $~\alpha_s(Q^2_0)$ at the initial scale $Q^2_0=1$ GeV$^2$.
Thus, given the value of $~\alpha_s(Q^2_0)$ we have numerically solved 
the differential equation for $~\alpha_s(Q^2)$ for any $Q^2$ \cite{num_calc}. For $~\alpha_s(Q^2_0)$ we have used the value 0.48780 obtained by the MMHT'14 NLO QCD analysis \cite{MMHT14} of the world unpolarized data, which corresponds to $~\alpha_s(M^2_{z}) = 0.120$.
This value was chosen in order that the $Q^2$ evolution of the polarized PDFs would be consistent with the evolution of the unpolarized MMHT'14 PDFs which are used in the positivity constraints. 

\section{Results of Analysis}

In this section we will present and discuss the results of our new
NLO QCD fit to the present world data on polarized inclusive DIS
(\cite{EG1bp}, \cite{EG1bd}, \cite{HERMES}, \cite{EMC}-\cite{JLabn}). 
The data used (682 experimental points) cover the following kinematic
region: $\{0.004 \leq x \leq 0.75,~~1< Q^2 \leq 96.1$~GeV$^2\}$. Note that
for the CLAS/EG1b data a cut $W > 2$ GeV was imposed in order to exclude
the resonance region.

In order to study the effects on the polarized PDFs on the deviation of 
$a_8$ from its $SU(3)$ symmetric value, we have performed the following fits:

(a) Fit A:The data is fitted using the $SU(3)$ symmetric value 0.586 for $a_8$ (\ref{3FD}).

(b) Fit B:The data is fitted using $a_8=0.46$. This value corresponds to
the maximal reduction of $a_8$ presented in the literature and is the
value predicted in one of the models on $SU(3)$-breaking effects
\cite{Bass-Thomas}.

(c) Fit C:The data is fitted using $a_8$ as a free parameter.

Since  the isospin $SU(2)$ symmetry is considered as almost exact, 
we have used the very precisely measured value $g_A=1.270$  (see Eq. (\ref{ga})).

The numerical results of our NLO QCD Fit A to the present world $g_1$ data
set are presented in Tables I - III.
\begin{table}[ht]
\caption{\label{tab:table1}  Data used in our NLO QCD
analysis, the individual $\chi^2$ for each set
and the total $\chi^2$ of the Fit A.}
\begin{ruledtabular}
\begin{tabular}{ccccccc}
Experiment~~~~~~~~~~~~~~&~~~Process~~~~&$~~N_{data}~~~$&
$~~~~\chi^2$~~~~ \\ \hline
 EMC \cite{EMC}~~~~~~~~~~~~~~~~&   DIS(p)   &  ~10 & ~6.6 \\
 SMC \cite{SMC}~~~~~~~~~~~~~~~~&    DIS(p)  &  ~12 & ~4.6  \\
 SLAC/E143 \cite{SLAC143}~~~~~~~& DIS(p) &  ~28 & 23.6 \\
 SLAC/E155 \cite{SLAC155p}~~~~~~~& DIS(p) & ~24 & 22.3 \\
 HERMES \cite{HERMES}~~~~~~~~~~& DIS(p)  & ~37 & 18.2 \\
 COMPASS'10 \cite{COMPASS10p}~~~~~& DIS(p) &  ~15 & 11.1 \\
 COMPASS'16 \cite{COMPASS16p}~~~~~& DIS(p) &  ~51 & 31.2 \\
 CLAS/EG1b \cite{EG1bp}~~~~~~~& DIS(p) & 166  &  91.0  \\ \\
 SMC \cite{SMC}~~~~~~~~~~~~~~~~~&    DIS(d)   &  ~12 & 17.2   \\
 SLAC/E143 \cite{SLAC143}~~~~~~~& DIS(d) &  ~28 & 41.4 \\
 SLAC/E155 \cite{SLAC155d}~~~~~~~& DIS(d) & ~24 & 17.9 \\
 HERMES \cite{HERMES}~~~~~~~~~~& DIS(d)  & ~37 & 35.7 \\
 COMPASS \cite{COMPASSd}~~~~~~~~& DIS(d) &  ~43 & 29.8 \\
 CLAS/EG1b \cite{EG1bd}~~~~~~& DIS(d) & 158  &  136.2  \\ \\
 SLAC/E142 \cite{SLAC142}~~~~~~& DIS(n) &  ~~8 & ~5.8 \\ 
 SLAC/E154 \cite{SLAC154}~~~~~~& DIS(n) &  ~17 & ~5.4 \\
 HERMES \cite{HERMESn}~~~~~~~~~~& DIS(n)  & ~~9 & ~2.6\\
 JLab-Hall A \cite{JLabn}~~~~~~& DIS(n)& ~~3& ~1.4 \\ \\

{\bf TOTAL}:~~~~~~~~~~~~~~&        &  {\bf 682} & {\bf 501.6}  \\
\end{tabular}
\end{ruledtabular}
\end{table}

In Table I the $g_1$ data sets used in our analysis are listed and the
corresponding values of $\chi^2$ obtained from the best fit to the
data are presented. As seen from Table I, a good description of
the data is achieved: $\chi^2/{d.o.f.}$=0.759 for 682 experimental
points using 21 free parameters (11 for the PDFs and 10 for the
higher twist corrections). The final proton and deuteron CLAS/EG1b data
are well consistent with the previous world data set.

The values of the parameters attached to the input polarized PDFs
obtained from the best fit to the data are presented in Table II.
The errors used in the fit are quadratic combinations of the statistical 
and point-to-point systematic errors. As seen from Table II, the parameters describing the polarized strange quark density are well determined. 
\begin{table*}
\caption{\label{tab:table2} The parameters of the NLO input
polarized PDFs at $Q^2=1$ GeV$^2$ obtained from the best Fit A to
the data. The errors shown are total (statistical and systematic).
The parameters marked by (*) are fixed.}
\begin{ruledtabular}
\begin{tabular}{ccccccc}
  flavor &  A  &  $\alpha$ &  $\beta$ & $\epsilon$ & $\gamma$  \\ \hline
 $u+\bar{u}$& ~1.9456$^*$ & 0.8947~$\pm$~0.1658 & 3.1975~$\pm$~0.1097 &
-1.3766~$\pm$~0.8079 & 5.2073~$\pm$~2.4380~ \\
 $d+\bar{d}$& -0.5062$^*$ & 0.5821~$\pm$~0.1256 & 4.2776~$\pm$~0.3998 &
0 & 5.0596~$\pm$~4.4773  \\
$s+\bar{s}$ & -0.2524~$\pm$~0.1663 & 0.5624~$\pm$~0.1661 & 9.0$^*$ &
0 & 0 \\
G & 30.241~$\pm$~13.384 & 2.9128~$\pm$~0.7419 & 5.0$^*$ & 0 &
0 \\
\end{tabular}
\end{ruledtabular}
\end{table*}

The extracted polarized NLO PDFs are plotted in Fig. 1 for
$Q^2=2.5$ GeV$^2$ and compared to those obtained by the groups
(\cite{BB}-\cite{SKAO18}). As seen from Fig. 1, our strange quark density 
is negative and consistent with that obtained in the previous analyses 
in which for $a_8$ its $SU(3)$ value was used. It is also seen that 
as a function of $x$, the shape of the JAM'15 strange quark density is 
harder than those obtained from the other groups. Note that contrary to 
the other groups the JAM Collaboration has used in his analysis \cite{JAM15} an alternative approach, based on a new iterative Monte Carlo fitting technique.
\begin{widetext}
\begin{center}
\begin{figure} [h]
  \includegraphics[scale=0.75]{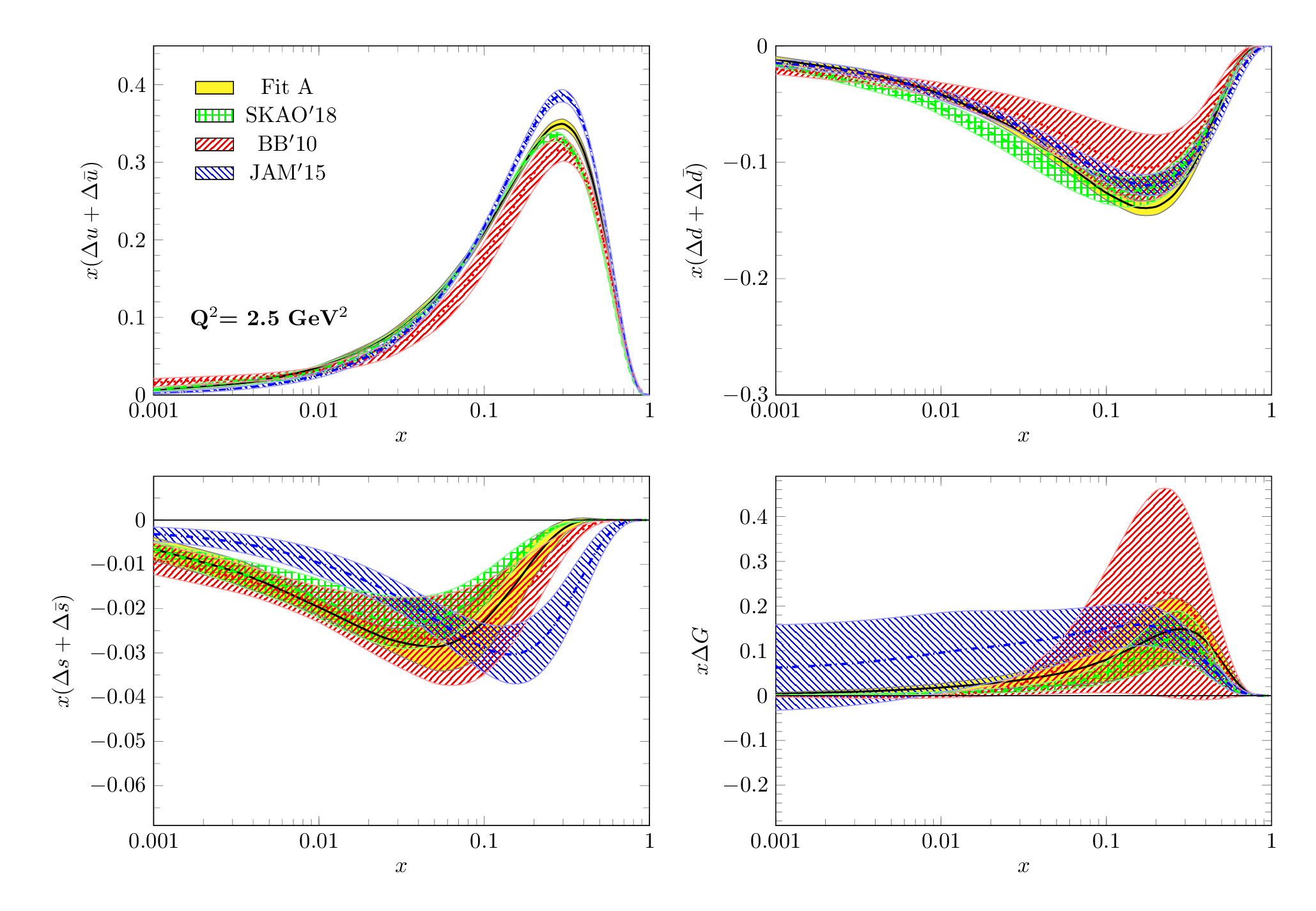}
\caption{Our NLO polarized PDFs obtained from Fit A($\gamma_{s_{+}}=0$) compared to those of BB'10 \cite{BB}, JAM'15 \cite{JAM15}, and SKAO'18 \cite{SKAO18}. 
\label{fig1} }
\end{figure}
\end{center}
\end{widetext}

Note that the results for the polarized PDFs presented in the Table I 
and Fig. 1 correspond to Fit A using $\gamma_{s_{+}}=0$ 
in the input parametrization for the polarized strange quark density (\ref{input_PDFs}). As was shown in \cite{LSS15} the precision of the
the data available at that time was enough in order to determine well  
the parameter $\gamma_{s_{+}}$. This parameter is important in principle
because it allows changing in sign behaviour for the polarized strange 
quark density. So, we have repeated Fit A using $\gamma_{s_{+}}$ as a 
free parameter, and obtained for it: $\gamma_{s_{+}}=-2.859 \pm 0.552$.
The values obtained for $\chi^2/{d.o.f.}$ and the first moments of the 
quark densities, are essentially the same as in the case $\gamma_{s_{+}}=0$. 
So, including in the data fit one more parameter does not improve the description of the data. Although the first moment
of $(\Delta s+\Delta \bar{s})(x)$ corresponding to $(\gamma_{s_{+}} \neq 0)$
is equal to that with $(\gamma_{s_{+}} = 0)$, the behavior of the strange quark density is slightly different (see Fig. 2). 
\begin{figure}[bht]
\includegraphics[scale=0.80]{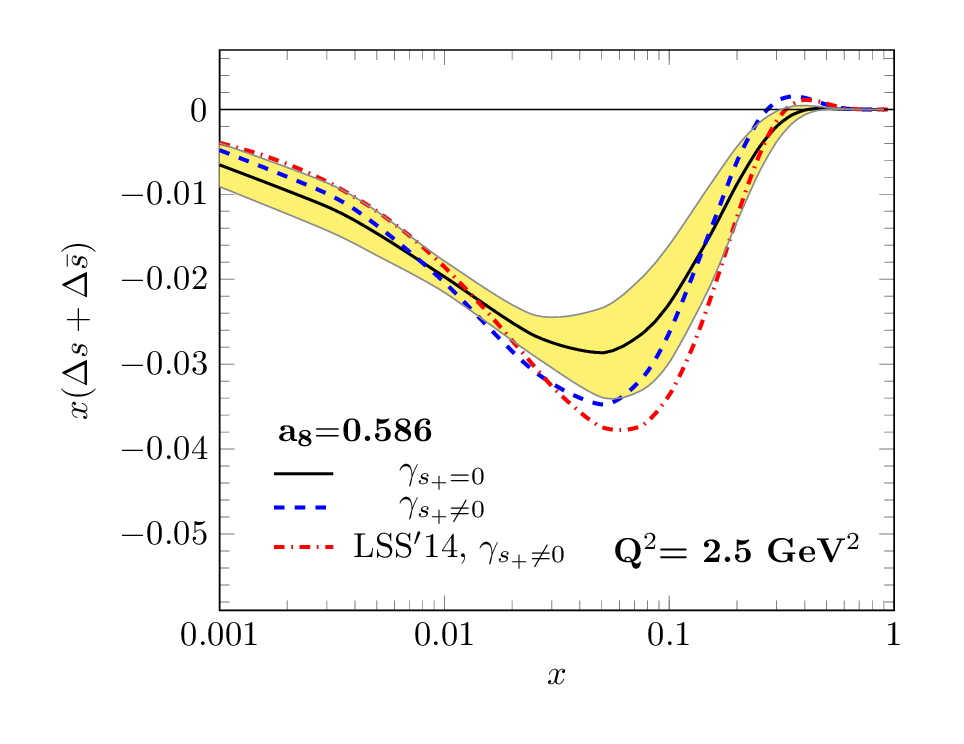}
\caption{Comparison between the strange quark densities obtained  
in the Fit A with $\gamma_{s_{+}}=0$ and $\gamma_{s_{+}} \neq 0$
(see the text). The LSS'14 strange quark density \cite{LSS15} ($\gamma_{s_{+}} \neq 0$) is also shown. 
\label{dels} }
\end{figure}

It is important to note that the strange quark density is {\it negative} for small values of $x$ and only changes sign in the region $0.2<x< 0.35$ (the precise point depending on the value of $Q^2$). Beyond this cross-over point it is exceedingly small, compatible with zero (see Fig. 2). As seen from Fig. 2 it is consistent with the LSS'14 strange quark density \cite{LSS15}. The difference in the large $x$ region is due to the different behavior of the unpolarized strange qurk densities used in the positivity constraints, the MRST'02 set in \cite{LSS15} and MMHT'14 in this analysis.

\begin{table*}
\caption{\label{tab:table3} Sensitivity of the first moments of the polarized parton densities to $SU(3)$ symmetry flavour symmetry breaking 
($Q^2 = 2.5$ GeV$^2$). The $SU(3)$ value 3F-D=0.586.}
\begin{ruledtabular}
\begin{tabular}{ccccccc}
 ~~$a_8$~~&$\chi^2/{d.o.f}$&$(\Delta u + \Delta\bar{u})$ &
 $(\Delta d + \Delta\bar{d})$ & $(\Delta s + \Delta\bar{s})$ 
&$\Delta \Sigma$&$\Delta G$\\ \hline
 3F-D (Fit A)& 0.759 & 0.815~$\pm$~0.025 & -0.455~$\pm$~0.030 & -0.113~$\pm$~0.020 & 0.247~$\pm$~0.044 & 0.328~$\pm$~0.161\\
 0.46 (Fit B) & 0.759 & 0.805~$\pm$~0.024 & -0.465~$\pm$~0.029 & -0.060~$\pm$~0.016 & 0.280~$\pm$~0.041 & 0.331~$\pm$~0.160\\ 
\end{tabular}
\end{ruledtabular}
\end{table*}

\begin{figure}[bht]
\includegraphics[scale=0.75]{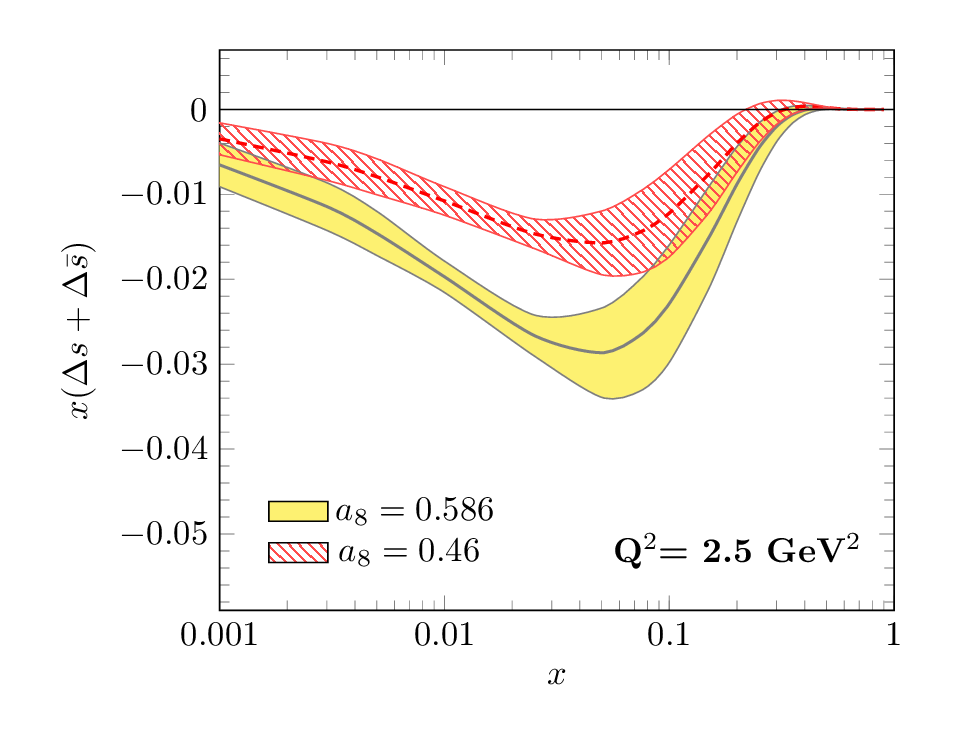}
\includegraphics[scale=0.75]{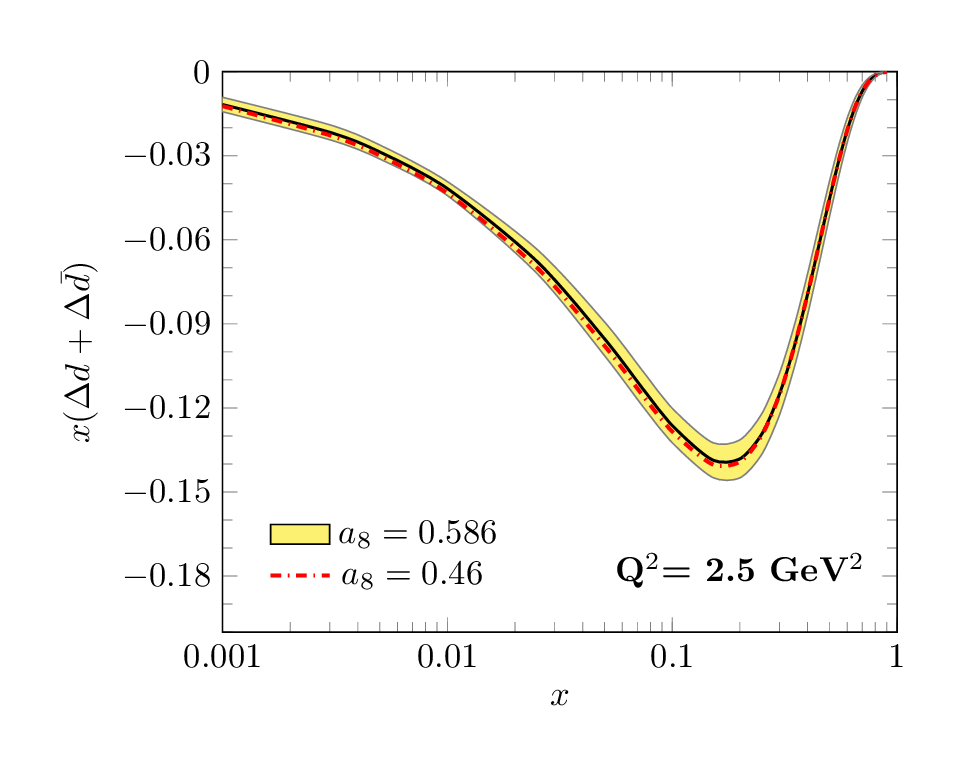}
\caption{Comparison between the polarized strange quark densities (top)
and ($\Delta d + \Delta \bar{d}$) densities (bottom) obtained in the Fit A using for $a_8$ its $SU(3)$ value and Fit B $(a_8=0.46)$.}
\label{dels_diffr_a8}
\end{figure}

Let us comment now the results of Fit B when for the nonsinglet 
axial charge $a_8$ instead of its $SU(3)$ symmetric value 0.586, 
the value 0.46 have been used in the fit. In order to demonstrate the sensitivity of the parton densities to the $SU(3)$ breaking we present 
them (Fig. 3) and the values of their first moments in Table III, and compare them with those obtained in the Fit A. One can see from Table III that the values of $\chi^2/d.o.f.$ for both the fits are the same, 
so that the present polarized inclusive DIS data cannot distinguish between these two values. Note that all results of Fit B presented 
in the paper correspond to the parameter $\gamma_{s_{+}}=0$. The usage of  
$\gamma_{s_{+}}$ as a free parameter does not essentially change the 
results and conclusions.

Contrary to the rest of the parton densities, which are essentially those determined by the $SU(3)$ analysis of the data, the polarized strange quark density changes significantly when flavour $SU(3)$ symmetry is broken (see Fig. 3 and Table III). Compared to the $SU(3)$ case the shape of the strange quark density is almost the same, but its magnitude is 
approximately halved. It is important to note that with the inclusion of much new data the strange quark density and its first moment remain significantly negative, in agreement with the result obtained by LSS group from the analysis of the polarised inclusive DIS data available at that time (see the footnote [27] in \cite{LSS15}). 

As seen from Table III, as a result of the reduction of the $SU(3)$ value of $a_8$ by $22 \%$, causes the first moment of the singlet 
quark density, $\Delta \Sigma$ (the spin of the nucleon carried by the quarks), to increase by $13 \%$. 

In Fig. 3, in addition to the polarized strange quark density, we show
the fit B ($\Delta d + \Delta \bar{d}$) parton density as an 
illustration that the rest of the polarized parton densities are almost identical to those obtained in the $SU(3)$ Fit A.

As usual the polarized gluon density is extracted from the inclusive 
DIS data with a larger uncertainty than that for the other densities. 
Nevertheless, we would like to mention that the value of the truncated 
first moment of our gluon density, 
$\int_{0.05}^{1}dx\Delta G (x) =0.28 \pm 0.12$ at $Q^2=10$ GeV$^2$, is well 
consistent with that, $0.20 \pm 0.06$, determined from a global QCD analysis of the polarized parton densities \cite{delG_Vogelsang} including the high-statistics RHIC data on the double-spin asymmetries for inclusive jet and $\pi^{0}$ production \cite{RHICdata}. 

Finally, we will briefly mention our results for the fit to the data using $a_8$ as a free parameter (Fit C). For $\chi^2/d.o.f.$ we find the value
0.760 (practically the same value as for the Fits A and B, 0.759),
and for $a_8 = 0.322 \pm 0.018$, which implies a 45 \% violation of the $SU(3)$ flavor symmetry. However, such a strong violation of the $SU(3)$ flavor symmetry is in a big contradiction with the data on hyperon $\beta$ decays \cite{Cabibo}. The strange quark density is presented in Fig. 4 and compared to those obtained in Fit A and Fit B. As seen from Fig. 4, it is consistent with zero for all $x$ values, and its uncertainty is larger than that obtained in Fit A and fit B. 

In Fig. 4 is also shown the strange quark density (NNPDFpol1.0) obtained in the QCD analysis of the NNPDF Collaboration \cite{NNPDF} of the inclusive DIS data, in which the flavor $SU(3)$ symmetry has been broken in another way. For the nonsinglet charge $a_8$ has been  used the value $a_8= 0.585 \pm 0.176$, i.e. for the uncertainty has been used a value approximately six times bigger than the value 0.03 obtained from the phenomenological analysis of the data on hyperon $\beta$ decays. This choice of the uncertainty actually means that the $SU(3)$ symmetry is up to 30 \% broken. As a result,
the NNPDF central value for the strange quark density is close to ours 
(Fit A), however, their error band is much larger than that obtained in our Fit A, which is demonstrated in Fig. 4. (Note that our uncertainty for the Fit A strange quark density is close to those obtained in the fits using the value 0.03 for the uncertainty of $a_8$, see Fig. 1.)

We would like to mention also that in this analysis a $W^2$ cut of the data has been used in order to minimize the higher twist effects. However, such a cut of the data in the polarized case,    
results in the loss of information from the already much smaller amount of data than in the unpolarized case. Note that the uncertainty for the strange quark density is slightly improved in the small x region, $x < 0.001$ from a combined analysis of the same collaboration \cite{NNPDF'14}, where to the polarized inclusive DIS data the polarized hadron collider data for inclusive jet and W boson production from the STAR and PHENIX experiments at RHIC have been added. 
\begin{figure}[bht]
\includegraphics[scale=0.80]{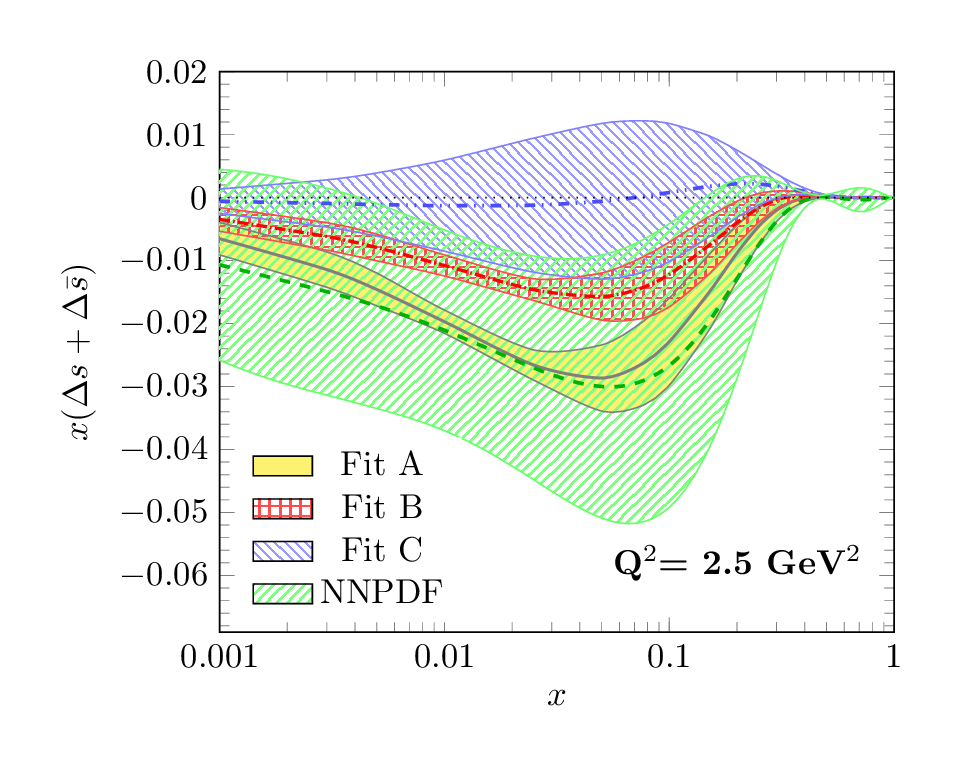}
\caption{Comparison between the strange quark densities obtained from
Fits A, B, C, and NNPDFpol1.0\cite{NNPDF}}
\label{fig4}
\end{figure}

The JAM Collaboration, based on a global QCD analysis of polarized inclusive 
and semi-inclusive deep-inelastic scattering and single-inclusive $e^{+}e^{-}$ annihilation data using $a_8$ as a free parameter \cite{JAM17}, found the value 0.46 which we used in our Fit B.
Thus in Fig. 5  we compare the polarized strange quark density obtained from fit B with the JAM'17 result.
In the JAM'17 analysis the polarized parton densities and the fragmentation functions have been simultaneously extracted from the data, but the very precise JLab inclusive DIS data were not included in the fit. Also, the fragmentation functions are mainly fixed from the
semi-inclusive data on the longitudinal double spin asymmetries, which 
are much less precise than the unpolarzed semi-inlusive data on hadron multiplicities, the best source for their precise determination. As a result, the uncertainties for the polarized parton densities are much larger, especially for the strange quark density, than those obtained from the inclusive DIS data. As seen from Fig. 5, our Fit B polarized  strange quark density together with its error band, entirely lies within the large error band of the JAM'17 strange quark density. 
\begin{figure}[bht]
\includegraphics[scale=0.80]{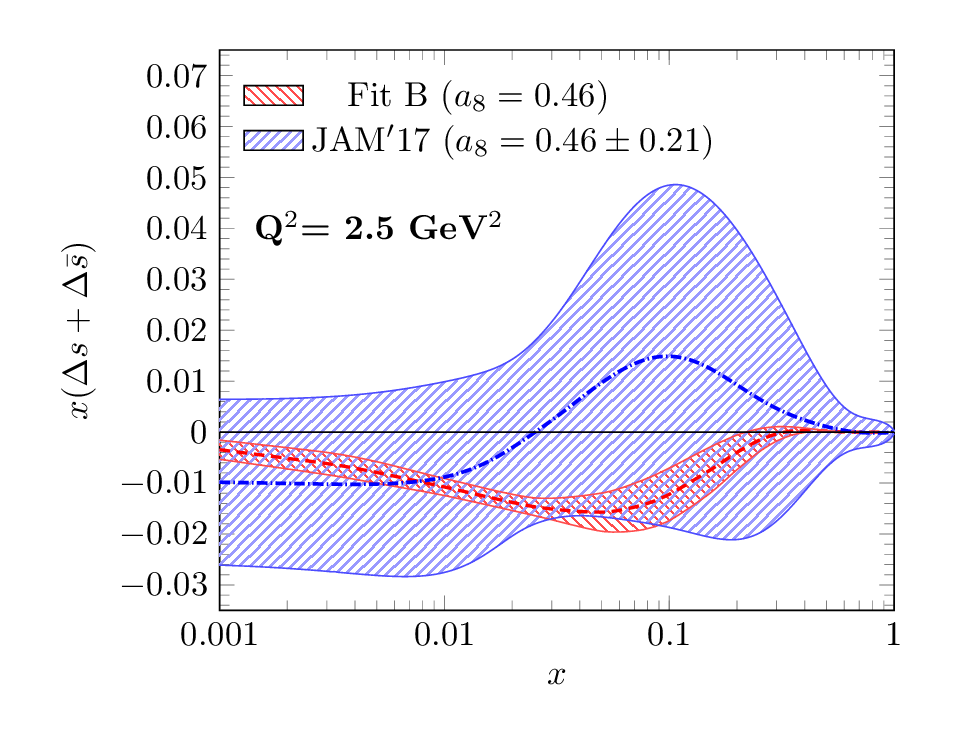}
\caption{Comparison between the strange quark densities obtained  
in the Fit B ($a_8=0.46$) and JAM'17 \cite{JAM17}.}
\label{dels_a_8red_JAM17} 
\end{figure}

Interestingly, the JAM'17 value 0.46 for $a_8$ agrees with the value predicted in \cite{Bass-Thomas}. However, bearing in mind the large error, 45\% of the magnitude of $a_8$, we are still 
far from really fixing the magnitude of the breakdown of the $SU(3)$ flavor 
symmetry from the polarized DIS and SIDIS data.

We would like to stress once more that in order to obtain reliable values for the polarized individual quark and antiquark densities, as well as for $a_8$, using SIDIS data, it is crucial that the FFs to be well determined. In this context it should be noted that the extracted pion and kaon FFs in the JAM'17 analysis are not consistent with those obtained from the global analysis \cite{DSS+}, where the unpolarized SIDIS data on hadron multiplicities were used.    
\begin{figure}[ht]
\includegraphics[scale=0.75]{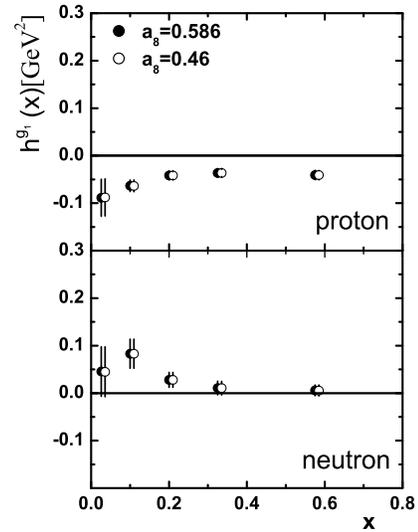}
\caption{Comparison between higher twist corrections corresponding to 
the values of $a_8$ 0.586 and 0.46. }
\label{fig6}
\end{figure}

As was mentioned above (see Eq.~ (\ref{HTQCD})), we have taken into account the higher twist corrections $h^N(x_i)/Q^2$ to the spin structure functions in our fits to DIS data, treating the $h^N(x_i)$ as are free parameters. 
The values
of the HT corrections $h^p(x_i)$ and $h^n(x_i)$ for the proton and
neutron targets extracted from the data in this analysis are
presented in Fig. 6. For the deuteron target the relation
$h^d(x_i)=0.925[h^p(x_i)+h^n(x_i)]/2$ has been used, where 0.925
is the value of the polarization factor $D$.

As seen from Fig. 6, the extracted from the data higher twist corrections
are not sensitive to the value of nonsinglet axial charge $a_8$. 

\section{Summary}

We have presented a NLO QCD analysis of the present world inclusive DIS data 
on the nucleon spin structure functions $g_1^N$, in which the sensitivity of 
the polarized strange quark density to  flavor $SU(3)$ symmetry breaking, 
i.e. to the deviation of the value of the non-singlet axial charge $a_8$ 
from its $SU(3)$ symmetric value 0.586, has been studied. Three fits to the data corresponding to different values of $a_8$ were performed using: 
(A) its $SU(3)$ symmetric value 0.586,  (B) the value 0.46 i.e. the maximal 
reduction of $a_8$ presented in the literature and obtained in a  theoretical model, and (C) $a_8$ taken as a free parameter to be obtained from the best fit to the data.

It was shown that contrary to the rest of the parton densities, which 
hardly change from their $SU(3)$ analysis values, the polarized strange quark density changes significantly when flavour $SU(3)$ symmetry is broken. When $a_8=0.46$, the strange quark density and its first moment still remain significantly negative. Compared to the $SU(3)$ case, 
the shape is almost the same, but its magnitude is approximately halved. 
Using $a_8$ as a free parameter we obtain $a_8 = 0.322 \pm 0.018$ and in this case the strange quark density is consistent with zero. 
The above value for $a_8$ implies a 45 \% violation of  $SU(3)$
flavor symmetry which is unlikely to be consistent with the data 
on hyperon $\beta$ decays. 

An important  feature of all the fits is that the data are well described 
and the value of $\chi^2/{d.o.f.}$ is practically the same for all three types of fit, which means that the inclusive  polarized DIS data alone cannot distinguish between the different strange quark densities discussed above. 

To improve the sensitivity to $SU(3)$ breaking and to the polarized strange 
quark density, at the very least, polarized semi-inclusive DIS data have 
to be involved, and such attempts have been made in the past 
(\cite{DSSV}-\cite{LSS11}) and very recently \cite{JAM17}. However, their success depends on how reliably the pion and kaon fragmentation functions used in the analysis are determined. Bearing in mind that the very precise HERMES and COMPASS experimental data on the pion and kaon multiplicities are inconsistent \cite{COMPASS_mult}, it seems that the carrying out of such an analysis is, at present, not so easy. 
In the long run data from an Electron-Ion Collider, with a much larger range of $Q^2$, will help to constraint much better the strange quark polarisation in the nucleon, but it is essential that any such results should correspond to the extracted value for $a_8$  being consistent with the data on hyperon $\beta$ decays. \\

\begin{center}
{\bf ACNOWLEDGMENTS}
\end{center}

E. Leader and D. Stamenov are grateful to W. Melnitchouk for
providing us with the JAM'15 and JAM'17 polarized PDFs, as well as for the useful discussion. A.K. appreciate A. Vogt and H. Abdolmaleki for discussions and guidance. This research was supported by Semnan University.

\end{document}